\documentclass[sigconf]{acmart}
\usepackage{balance}
\usepackage{graphicx}
\usepackage{epsfig}
\usepackage{caption}   % 加载 caption 包
\usepackage{amsmath}
\usepackage{svg}
\usepackage{booktabs}
\usepackage{multirow}
\usepackage[normalem]{ulem}
\useunder{\uline}{\ul}{}
\usepackage{enumerate}
\usepackage{bbding}
\AtBeginDocument{%
  }

\copyrightyear{2025}
\acmYear{2025}
\setcopyright{acmlicensed}\acmConference[KDD '25]{Proceedings of the 31st ACM
SIGKDD Conference on Knowledge Discovery and Data Mining V.2}{August 3--7,
2025}{Toronto, ON, Canada}
\acmBooktitle{Proceedings of the 31st ACM SIGKDD Conference on Knowledge
Discovery and Data Mining V.2 (KDD '25), August 3--7, 2025, Toronto, ON, Canada}
\acmDOI{10.1145/3711896.3737172}
\acmISBN{979-8-4007-1454-2/2025/08}

\begin{document}

%%
%% The "title" command has an optional parameter,
%% allowing the author to define a "short title" to be used in page headers.
\title{Unlocking the Power of Diffusion Models in Sequential Recommendation: A Simple and Effective Approach}

\author{Jialei Chen}
\orcid{0009-0003-9861-7788}
\affiliation{%
\institution{MIC Lab,\\College of Computer Science and Technology,\\Jilin University}
% \institution{Department of Computer Science and Technology}
\city{Changchun}
\country{China}
}
\email{chenjl21@mails.jlu.edu.cn}

\author{Yuanbo Xu}
\orcid{0000-0001-8370-5011}
\authornote{corresponding author}
\affiliation{%
\institution{MIC Lab,\\College of Computer Science and Technology,\\Jilin University}
\city{Changchun}
\country{China}
}
\email{yuanbox@jlu.edu.cn}

\author{Yiheng Jiang}
\orcid{0000-0001-6737-0389}
\affiliation{%
\institution{MIC Lab,\\College of Computer Science and Technology,\\Jilin University}
\city{Changchun}
\country{China}
}
\email{jiangyh22@mails.jlu.edu.cn}

\renewcommand{\shortauthors}{Jialei Chen, Yuanbo Xu, \& Yiheng Jiang}

\begin{abstract}
In this paper, we focus on the often-overlooked issue of embedding collapse in existing diffusion-based sequential recommendation models and propose ADRec, an innovative framework designed to mitigate this problem. Diverging from previous diffusion-based methods, ADRec applies an independent noise process to each token and performs diffusion across the entire target sequence during training. ADRec captures token interdependency through auto-regression while modeling per-token distributions through token-level diffusion. This dual approach enables the model to effectively capture both sequence dynamics and item representations, overcoming the limitations of existing methods. To further mitigate embedding collapse, we propose a three-stage training strategy: (1) pre-training the embedding weights, (2) aligning these weights with the ADRec backbone, and (3) fine-tuning the model. During inference, ADRec applies the denoising process only to the last token, ensuring that the meaningful patterns in historical interactions are preserved. Our comprehensive empirical evaluation across six datasets underscores the effectiveness of ADRec in enhancing both the accuracy and efficiency of diffusion-based sequential recommendation systems.
% \href{https://github.com/Nemo-1024/ADRec}{The ADRec code is available at \textit{https://github.com/Nemo-1024/ADRec}.}
\footnotetext[2]{The ADRec code is available at \textit{https://github.com/Nemo-1024/ADRec}.}

\end{abstract}


%%
%% The code below is generated by the tool at http://dl.acm.org/ccs.cfm.
%% Please copy and paste the code instead of the example below.
%%
% \begin{CCSXML}
% <ccs2012>
%  <concept>
%   <concept_id>00000000.0000000.0000000</concept_id>
%   <concept_desc>Do Not Use This Code, Generate the Correct Terms for Your Paper</concept_desc>
%   <concept_significance>500</concept_significance>
%  </concept>
%  <concept>
%   <concept_id>00000000.00000000.00000000</concept_id>
%   <concept_desc>Do Not Use This Code, Generate the Correct Terms for Your Paper</concept_desc>
%   <concept_significance>300</concept_significance>
%  </concept>
%  <concept>
%   <concept_id>00000000.00000000.00000000</concept_id>
%   <concept_desc>Do Not Use This Code, Generate the Correct Terms for Your Paper</concept_desc>
%   <concept_significance>100</concept_significance>
%  </concept>
%  <concept>
%   <concept_id>00000000.00000000.00000000</concept_id>
%   <concept_desc>Do Not Use This Code, Generate the Correct Terms for Your Paper</concept_desc>
%   <concept_significance>100</concept_significance>
%  </concept>
% </ccs2012>
% \end{CCSXML}

% \ccsdesc[500]{Do Not Use This Code~Generate the Correct Terms for Your Paper}
% \ccsdesc[300]{Do Not Use This Code~Generate the Correct Terms for Your Paper}
% \ccsdesc{Do Not Use This Code~Generate the Correct Terms for Your Paper}
% \ccsdesc[100]{Do Not Use This Code~Generate the Correct Terms for Your Paper}

%%
%% Keywords. The author(s) should pick words that accurately describe
%% the work being presented. Separate the keywords with commas.
%% A "teaser" image appears between the author and affiliation
%% information and the body of the document, and typically spans the
%% page.

%%
%% This command processes the author and affiliation and title
%% information and builds the first part of the formatted document.

\maketitle
\newcommand\kddavailabilityurl{https://doi.org/10.5281/zenodo.15470542}

\ifdefempty{\kddavailabilityurl}{}{
\begingroup\small\noindent\raggedright\textbf{KDD Availability Link:}\\
% please change the following context to include multiple artifacts if necessary.
The source code of this paper has been made publicly available at \url{\kddavailabilityurl}.
\endgroup
}
% \begin{figure}[t]  % 'h' 表示将图片放置在此处
%     \centering  % 图片居中
%     \includesvg{img/test.svg}
%     \caption{1
%     } 
% \end{figure}

% Please add the following required packages to your document preamble:
% \usepackage{booktabs}
% \usepackage{graphicx}

\section{Introduction}
\label{sec:intro}

\begin{figure*}
% \vspace{-0.1cm}
    \centering
    \includegraphics[width=0.98\linewidth]{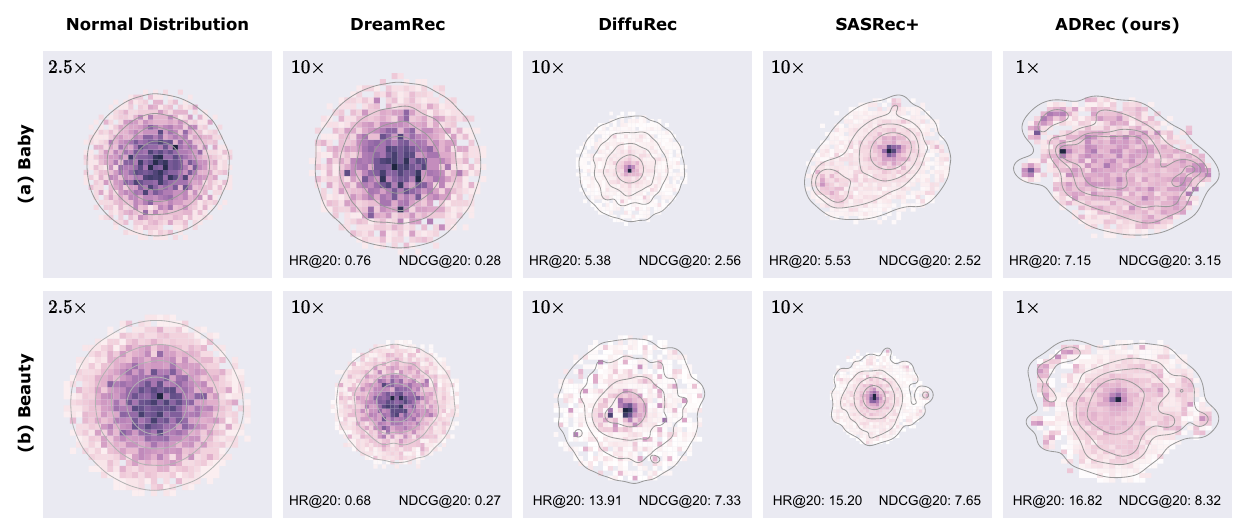}
    % \vspace{-0.2cm}
    \caption{T-SNE results of the learned item embeddings of ADRec and other baselines on the Baby and Beauty dataset. If the contour shape closely resembles isotropic Gaussian noise (as seen in DreamRec and DiffuRec) or if the representation space is narrow (as observed in DreamRec, DiffuRec, and SASRec+, which requires a large magnification factor), it suggests a weak embedding space. In contrast, ADRec maintains a structured embedding space and expands it compared to SASRec+, significantly enhancing item separability. Additional visualization results can be found in Appendix Figure \ref{fig:tnse_full}.\protect\footnotemark[1]}
    \label{fig:tsne}
\end{figure*}

Sequential recommendation (SR) has long been a cornerstone of modern recommendation systems. Its primary objective is to analyze historical interaction records between users and the system to predict the next item a user is likely to engage with. \citep{pan2024surveysequentialrecommendation, wang2019sequential}.
Various sequence models, such as Recurrent Neural Networks (RNNs) and Transformers \cite{vaswani2017attention}, have been widely adopted in SR. Transformer-based methods, which leverage the powerful self-attention mechanism \cite{vaswani2017attention}, including SASRec \cite{kang2018self} and BERT4Rec \cite{sun2019bert4rec}, have emerged as some of the most effective models for SR. 
Despite these advancements, sequential recommendation still faces significant challenges, such as weak representation spaces \cite{linSurveyDiffusionModels2024a, wang2023conditionaldenoisingdiffusionsequential, yangGenerateWhatYou2023}.

Diffusion models \cite{song2020score, ho2020denoising} may be regarded as a distinctive form of self-supervised learning due to their inherent diffusion process \cite{ddae2023, chen2024deconstructing, linSurveyDiffusionModels2024a, fuest2024diffusionmodelsrepresentationlearning}. They have exhibited notable capabilities in representation learning, which could prove invaluable in accurately capturing the distribution of item representations within recommendation systems. Nevertheless, their efficacy within recommendation systems has been comparatively disappointing in recent years, failing to yield the anticipated advancements. This elicits a critical inquiry: \textit{Are diffusion models fundamentally ill-suited for recommendation systems, or do they encounter intrinsic limitations in this domain?}

% We hypothesize that this issue stems from \textbf{embedding collapse}. As shown in Figure \ref{fig:tsne}, the Transformer-based SASRec+ \cite{klenitskiy2023turning} forms a structured yet narrow and fragile representation space, which is weak and lacks robustness, a characteristic also shared by other traditional recommendation models. Likewise, the representation spaces of diffusion-based models such as DiffuRec \cite{liDiffuRecDiffusionModel2023} and DreamRec \cite{yangGenerateWhatYou2023} are narrow and meaningless, which resemble isotropic normal distributions. We believe the representation spaces of DiffuRec and DreamRec have collapsed significantly. Embedding collapse greatly diminishes the model’s ability to distinguish between items, impeding its ability to predict user preferences accurately.

We hypothesize that the issue arises from \textbf{embedding collapse in diffusion-based methods}. As shown in Figure \ref{fig:tsne}, the Transformer-based SASRec+ \cite{klenitskiy2023turning} forms a structured yet narrow and fragile representation space, which lacks robustness—a limitation shared by other traditional recommendation models. Diffusion-based methods like DiffuRec \cite{liDiffuRecDiffusionModel2023} and DreamRec \cite{yangGenerateWhatYou2023} yield representation spaces that are not only narrow but also resemble isotropic Gaussian distributions, indicating a severe collapse. This resemblance is particularly concerning, as randomly initialized embeddings also follow such distributions, suggesting that these models fail to learn meaningful representations beyond their initial state.

To tackle the embedding collapse, we first summarize the architectural design and training strategies of existing methods in Table \ref{tb: comparison} and point out several potential flaws that could impact the use of diffusion models in sequential recommendation:

\begin{enumerate}[0]
    \item[$\bullet$] \textbf{Only performing denoising learning on the last item is insufficient.} Existing methods, including DiffuRec, DimeRec \cite{liDimeRecUnifiedFramework2024}, and DreamRec, focus on diffusion and compute loss only for the final target item instead of for the entire one-position offset target sequence. As a result, only a limited number of items are subjected to diffusion, significantly complicating the process of learning item distributions and contributing to embedding collapse. From an auto-regressive perspective, they miss the chance to implement per-token teacher forcing, which restricts the effectiveness of sequence modeling.
    
    \item[$\bullet$] \textbf{Inappropriate training loss impacts the effectiveness of diffusion models.} Some works, like DreamRec, use only denoising loss (Mean Squared Error, MSE) as the training objective. This method does not align with the recommendation task and often produces suboptimal results. DiffuRec \cite{liDiffuRecDiffusionModel2023} retains recommendation loss (Cross-Entropy, CE) as the training objective but does not incorporate denoising loss, which limits the ability to exploit the potential of diffusion models fully.  
    \footnotetext[1]{Embeddings are normalized before T-SNE to ensure that the scale of the visualization results is comparable. "$10\times$" indicates that the visualization results have been magnified tenfold. In this context, "Normal Distribution" serves as a baseline, representing randomly initialized embedding weights.}

    \item[$\bullet$] \textbf{End-to-end training leads to Embedding collapse.} In the original application scenarios of diffusion models, input features (from images, audio, etc.) are fixed, meaningful, and distinguishable from one another. Item embeddings initialized from scratch are random and lack significance \cite{gao2024empoweringdiffusionmodelsembedding}, which creates a risk of embedding collapse, where all embedding weights converge to the same value to make the denoising learning easier. This collapse can cause the denoising loss to become zero during training, leading to poor performance during the inference phase. 

    \item[$\bullet$] \textbf{The sequence-level diffusion process causes a mismatch during the inference phase.} It affects the model’s inference by introducing noise throughout the entire sequence, which can corrupt the historical sequence during inference. This, in turn, impacts the quality of the target item prediction during inference.
\end{enumerate}

To tackle these challenges, we introduce \textbf{ADRec} (\textbf{A}uto-regressive \textbf{D}iffusion \textbf{Rec}ommendation model), a novel approach that merges a token-level diffusion process with causal attention for denoising. Unlike earlier diffusion-based methods, ADRec applies independent noise levels to each token within the entire one-position offset target sequence. This method guarantees that every token in the sequence undergoes diffusion, enabling the model to better learn item distributions while maintaining the per-token teacher-forcing feature of auto-regression. By capturing item dependencies through auto-regression and concurrently learning the distribution of each item via diffusion, ADRec can more effectively model both sequence dynamics and item representations.

To further mitigate embedding collapse, we introduce a three-stage training strategy. In the first stage, we pre-train the embedding weights using a causal attention module that helps ensure the embedding space is structured before full parameter training begins. During this stage, the model is similar to SASRec+ and employs CE loss. In the second stage, we warm up the backbone of ADRec to align its parameters with the pre-trained embedding weights, freezing the embedding weights during this phase. Finally, in the third stage, we conduct full-parameter training of ADRec. The last two stages employ a combination of CE and MSE to align the recommendation task while fully utilizing the distribution modeling capabilities of the diffusion process to optimize the embedding weights, ultimately creating a robust and structured embedding space (see last column in Figure \ref{fig:tsne}).

During the inference phase, the token-level diffusion process applies denoising iterations solely to the last target while assigning the diffusion time steps for the other positions to zero. The independent noise process prevents the introduction of unreasonable noise into historical interactions, enabling ADRec to concentrate on meaningful patterns within the sequence.

To this end, ADRec overcomes all the previously mentioned limitations. We empirically demonstrate its effectiveness across six datasets of varying scales. The experimental results show that ADRec significantly outperforms mainstream baselines, achieving improvements of 15.45\% and 13.02\% on HR@20 and NDCG@20, respectively. Despite employing a three-stage training framework, ADRec reduces the training time by an average of 70.98\% compared with the best diffusion baseline.

\begin{table*}[t]
\vspace{-0.2cm}
\caption{A comparison of the architectural design and training strategies between ADRec and prior works.}
\label{tb: comparison}
\vspace{-0.2cm}
\centering
\resizebox{0.98\linewidth}{!}{%
\begin{tabular}{@{}cccccccc@{}}
\toprule
      & Diffusion Modelling & Loss Computation   & Train Objective   & Denoising Model           & Noise Process        & Embed.Pre-train & \textbf{Embed.Collapse} \\ \midrule
DiffRec\cite{wangDiffusionRecommenderModel2023} (SIGIR '23)  & full sequence   & \textbf{per token} & MSE               & MLP                       & sequence-level & \XSolidBrush          & \Checkmark \\
DreamRec\cite{yangGenerateWhatYou2023} (NeurIPS '23) & only last token & only last token    & MSE               & MLP                       & sequence-level & \XSolidBrush          &\Checkmark \\
DiffuRec\cite{liDiffuRecDiffusionModel2023} (TOIS '23) & only last token & only last token    & CE                & \textbf{Causal Attention} & sequence level & \XSolidBrush          & \Checkmark \\
DimeRec\cite{liDimeRecUnifiedFramework2024} (WSDM '24)  & only last token & only last token    & \textbf{CE + MSE} & MLP                       & sequence-level & \CheckmarkBold & \Checkmark\kern-1.2ex\raisebox{1ex}{\rotatebox[origin=c]{125}{\textbf{--}}}  \\
ADRec (ours) & \textbf{per token}  & \textbf{per token} & \textbf{CE + MSE} & \textbf{Causal Attention} & \textbf{token-level} & \CheckmarkBold     & \XSolidBold              \\ \bottomrule
\end{tabular}%
}
\end{table*}

\begin{figure*}[t]  % 'h' 表示将图片放置在此处
    \centering  % 图片居中
    \vspace{-0.2cm}
    \includegraphics[width=0.9\textwidth]{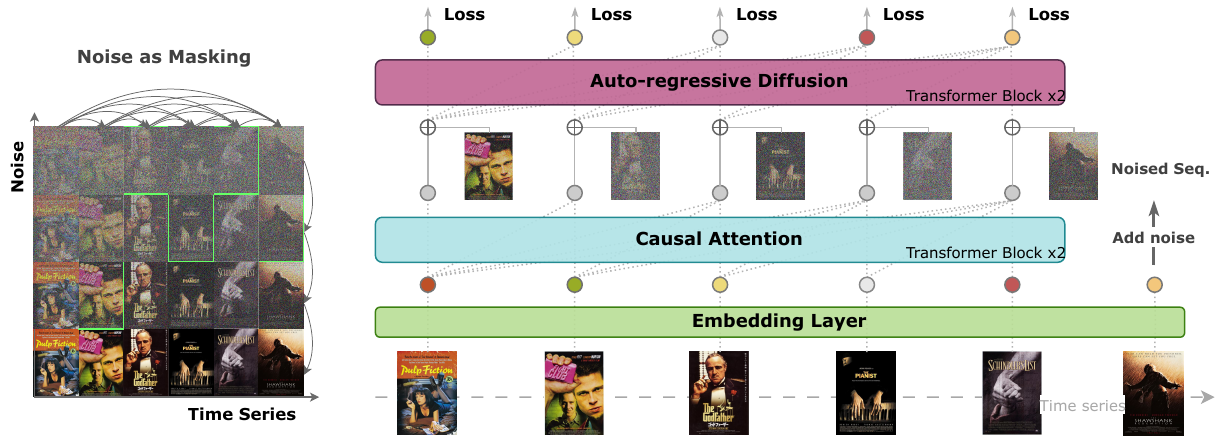}  % 图片路径
    \vspace{-0.2cm}
    \caption{Method overview. The left diagram illustrates that time series and diffusion processes are two orthogonal directions of evolution, with noise acting as a soft mask to measure uncertainty. ADRec applies independent diffusion processes to individual items, with the noise level of the items in the current target sequence highlighted in green box.\protect\footnotemark[2]
    } 
    \label{fig: overview}  % 图片标签，用于引用
\end{figure*}

\section{Background}
We have previously discussed several SR methods, and because of space limitations, we have included the Related Works in Appendix Section \ref{sec: apd related work}. Before presenting our model, we briefly introduce the standard diffusion model and sequential recommendation as foundational knowledge. 
\subsection{Diffusion Models}
Diffusion models \cite{ho2020denoising, song2020score} have gained popularity as generative modeling techniques in various domains. They can be seen as a specific method of self-supervised learning that utilizes a diffusion process to enhance representation learning.

\textbf{Forward Process} In this paper, we denote the diffusion process with the subscript $t$ to indicate diffusion time steps. For each token $x$, we define a forward diffusion process that progressively adds Gaussian noise to the data over a series of time steps. This process is modeled as a Markov chain, where the data at each step $k$ is incrementally noised.
\begin{equation}
    q\left(x_{t} \mid x_{t-1}\right)= \mathcal{N}\left(\sqrt{1-\beta_t} x_{t-1},\beta_t I\right)
    \label{eq:qx}
\end{equation}
where $\mathcal{N}$ is the normal distribution, and $\beta_t$ is the variance of the noise added at each step controlled by a schedule $\left\{\beta_{k} \in(0,1)\right\}_{k=1}^{K}$. The process continues until the data is converted into pure noise at $x_T$. With $\bar{\alpha}_t=\prod_{t^{^{\prime}}\leq t}(1-\beta_{t^{^{\prime}}})$, we can analytically write the result of the forward process given an original data:
\begin{equation}
\begin{aligned}
    q(x_t\mid x_0)&=\mathcal{N}(x_t;\sqrt{\bar{\alpha}_t}x_0,(1-\bar{\alpha}_t)I) \\
    i.e.\quad x_t&=\sqrt{\bar{\alpha}_t}x^0+\sqrt{(1-\bar{\alpha}_t)}\epsilon,\quad\epsilon\sim\mathcal{N}(0,I)
\end{aligned}
\end{equation}
$\bar{\alpha}_t$ is the noise schedule employed for the forward process. In original DDPM \cite{ho2020denoising}, $t$ is randomly sampled from a uniform distribution $\mathcal{U}[0,1]$ and a function maps $t$ to $\bar{\alpha}_t\in[0, 1]$.

\textbf{Reverse Process} The reverse process is also a Markov chain and attempts to recreate the original data from the noise with a denoising model $p_{\theta}(x_{t-1}\mid x_t)$:
\begin{equation}
    p_{\theta}\left(x_{t-1} \mid x_{t}\right)=\mathcal{N}\left(x_{t-1} ; \boldsymbol{\mu}_\theta\left(x_{t}, t\right), \gamma_{t}I\right)
\end{equation}
where the mean $\boldsymbol{\mu}_\theta$ is modeled with a neural network, and one can set the covariance to the identity scaled by a fixed constant depending on $t$. The $\boldsymbol{\mu}_\theta$ can be formulated by either using the noise $\epsilon$ or the target $x_0$:
\begin{align}
\mu_{\theta} & =\frac{1}{\sqrt{1-\beta_t}}x_t-\frac{\beta_t}{\sqrt{(1-\beta_t)(1-\bar{\alpha}_t)}}\epsilon  \label{eq: pre eps} \\
 & =\frac{\sqrt{1-\beta_t}(1-\bar{\alpha}_t))}{1-\bar{\alpha}_t)}x_t+\frac{\sqrt{\bar{\alpha}_{t-1})}\beta_t}{1-\bar{\alpha}_t}x_0
\label{eq: pre x}  
\end{align}
In Equation \eqref{eq: pre eps}, the model learns to predict the noise $\hat{\epsilon}$; while in Equation \eqref{eq: pre x}, the model learns to predict the original data $\hat{x}_0$. We use the latter formulation. Thus, a denoising model $f_\theta(x_t, t)$ can be trained to predict the original data $x_t$ that is available at timestamp $t$. Efficient training of DMs is possible by optimizing the simplified MSE loss instead of the original variational lower bound (VLB) as
\begin{equation}
     {\mathcal{L}}_{\text{simple }} = {\mathbb{E}}_{t,{x}_{0},\epsilon }\left\lbrack  {\begin{Vmatrix}{x}_{0}  - {f}_{\theta }\left( {x}_{t},t\right) \end{Vmatrix}}_{2}^{2}\right\rbrack 
\end{equation}

\subsection{Sequential Recommendation}
% Given an interacted list $S=[i^0,i^1,\dots,i^{L-1},i^L]$. Here $i^k$ denotes the $k$-th interacted item. $L$ is the max sequence length in the train set $\mathbf{i} = i^{0:L-1}$. The goal of sequential recommendation is to generate a ranked list of items as the predicted candidates for the next item that will be preferred by the user. In the training phase, there are typically two strategies for dividing the historical sequence and the target sequence. The first one is leaving one out where the target sequence is $\mathbf{i}_{tgt} = i^{L}$. We call this train strategy \textit{one-step prediction}, denoted as $g_\theta(i^{0:L-1}) = i^{L}$. The other utilizes the one-position-offset sequence  as the target sequence $\mathbf{i}_{tgt} = i^{1:L}$ and performs per-token teacher forcing as $g_\theta(i^{k}) = i^{k+1}$ for $0 \le k< L$, called \textit{per-step prediction}. They share the same inference process.

Given an interaction sequence $S=[i^0,i^1,\dots,i^{L-1},i^L]$, where $i^k$ denotes the $k$-th interacted item and $L$ indicates the maximum sequence length in the training set, the aim of SR is to produce a ranked list of items as predicted candidates for the next item the user is likely to interact. There are generally two auto-regressive training strategies for sequential recommendation models.

The first strategy, referred to as \textit{one-step prediction}, leaves one item out, using 
$\mathbf{i}_{tgt} = i^{L}$ as the target sequence. The model learns the mapping $g_\theta(i^{0:L-1}) = i^{L}$, where the model predicts the next item in the sequence. The second strategy, called \textit{per-step prediction}, shifts the sequence by one position, using $\mathbf{i}_{tgt} = i^{1:L}$ as the target sequence. This strategy employs per-token teacher forcing, where the model predicts each item in the sequence based on the previous item: $g_\theta(i^{k}) = i^{k+1}$ for $0 \le k< L$. Both strategies share the same inference process.

\section{Methodology}

\subsection{Auto-regressive strategy}
\label{sec: AR strategy}
In the previous section, we examined several auto-regressive training strategies that are essential for model architecture, as they directly affect the tensor shape of the target sequence. Let the target embedding sequence be denoted as $\mathbf{x}$. For one-step prediction, $\mathbf{x} = e^{L} \in \mathbb{R}^{B \times 1 \times D}$, and for per-step prediction, $\mathbf{x} = e^{1:L} \in \mathbb{R}^{B \times L \times D}$, where $B$ is the batch size, $L$ is the sequence length, and $D$ is the embedding dimension. For simplicity, we ignore sequence padding in this discussion. We adopt a per-token prediction strategy, ensuring that each output token contributes to the loss calculation and thus benefits from the improved sequence mining capabilities provided by per-token teacher forcing.

Existing diffusion-based SR algorithms typically employ one-step prediction, where denoising learning is only applied to the final target item, $e^{L}$ \cite{liDiffuRecDiffusionModel2023, yangGenerateWhatYou2023, liDimeRecUnifiedFramework2024, wang2023conditionaldenoisingdiffusionsequential,du2023sequentialrecommendationdiffusionmodels}. This approach has two main limitations. First, one-step prediction undermines the model's capacity to capture correlations between tokens, as it loses the per-token teacher forcing. Second, the available samples for diffusion learning become excessively sparse. For example, in the MovieLens-100K (ML-100K) dataset, there are a total of 1,008 items but only 938 sequences, which means that only 938 samples are utilized for denoising learning in one-step prediction. From an individual standpoint, this sparsity inhibits the effective learning of item distributions. From a broader perspective, it fails to sufficiently represent the entire embedding space, resulting in embedding collapse.

DiffuRec attempts to tackle this issue by partitioning the original target sequence into subsequences based on temporal order, giving each item in the sequence the chance to model the distribution. However, this strategy considerably increases both the number of sequences and the training time required. Essentially, it is similar to performing per-step prediction but without utilizing the parallelization capabilities of the self-attention mechanism, resulting in significantly longer training times.

\footnotetext[2]{The movie posters in Figure \ref{fig: overview} are for illustrative purposes only; during actual training, ADRec only utilizes user interaction data, like movie ID.}

\subsection{ADRec Architecture}
This section provides a detailed description of ADRec's architectural design. ADRec employs a multi-layer Transformer encoder at its backbone, which has demonstrated strong performance and versatility in modeling sequential dependencies. The architecture consists of two key components: the causal attention module and the auto-regressive diffusion module, each built using two Transformer encoder layers to maximize generality and to simplify the design (right in Figure \ref{fig: overview}). \textit{ADRec models the interdependence of tokens by auto-regression, jointly with the per-token distribution by token-level diffusion.}

\subsubsection{Causal Attention Module}
As illustrated in Figure \ref{fig: overview}, we define an embedding sequence $\mathbf{e} \in \mathbb{R}^{B \times L \times D}$, where $B$ represents the batch size and $D$ denotes the hidden size, as the semantic encoding of the intrinsic latent aspects captured by the interacted sequence $\mathbf{i} \in \mathbb{R}^L$. The primary role of causal attention module (CAM), parameterized as $\mathbf{CAM}\phi(\cdot)$, is to extract historical sequence information $\mathbf{c} \in \mathbb{R}^{B \times L \times D}$ that has not yet been corrupted by noise and use it as conditional guidance for the Auto-regressive Diffusion Module (ADM) during denoising learning.
\begin{equation} \mathbf{c} = \mathbf{CAM}\phi(\mathbf{e}) \end{equation} Notably, we do not incorporate positional encoding. Experimental results (Appendix Table \ref{fig: PE}) indicate that including positional encoding leads to a slight decrease in performance, as further discussed in Appendix Section \ref{sec: apd PE}. 

\begin{figure*}[t]
    \centering

    \includegraphics[width=0.9\linewidth]{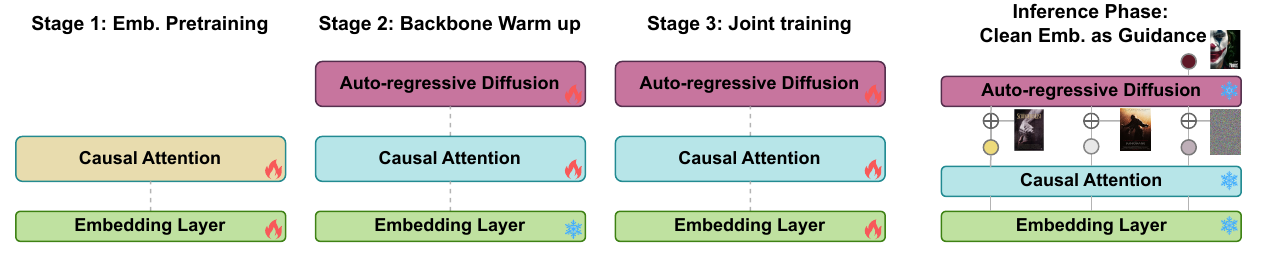}
    \vspace{-0.2cm}
    \caption{The three-stage training strategy and the inference strategy of ADRec.}
    \label{fig:stage}
\end{figure*}

\subsubsection{Feature Aggregation}
\label{sec: feat agg}
Before introducing the auto-regressive diffusion module, it is essential to define an appropriate feature aggregation method for the conditional guidance $\mathbf{c} \in \mathbb{R}^{B \times L \times D}$, the noised target sequence $\mathbf{x}_{t} \in \mathbb{R}^{B \times L \times D}$, and the current diffusion time step $\mathbf{t} \in \mathbb{R}^{B \times L}$.

We first scale the diffusion time step and encode it using an MLP with a SiLU activation function: \begin{equation} \mathbf{t}_{\textbf{emb}} = \mathbf{MLP}\left(\frac{1000 \cdot \mathbf{t}}{T}\right) \end{equation} Next, we aggregate the various components like DiffuRec \cite{liDiffuRecDiffusionModel2023} and denote the result as $\mathbf{z}$: 
\begin{equation} \mathbf{z} = \mathbf{c} + \lambda \odot (\mathbf{x}_t + \mathbf{t}_{\textbf{emb}}) \end{equation} 
where $\lambda$ is a coefficient, and we set $\lambda = 1e^{-3}$, which was the optimal setting in DiffuRec. We also experimented with replacing linear aggregation with cross-attention, but the results were unsatisfactory (see Figure \ref{fig: lambda} in Experiments).

\subsubsection{Auto-regressive Diffusion Module}

At this point, we introduce the auto-regressive diffusion module(ADM), parametrized as $\mathbf{ADM}_\theta(\cdot)$. The ADM module consists of a two-layer Transformer encoder, which also serves as the denoising model for the diffusion process. \begin{equation} \hat{\mathbf{x}} = \mathbf{ADM}_\theta(\mathbf{z}) \end{equation} where $\hat{\mathbf{x}}$ is the reconstructed target sequence.

\textbf{Token-independent Diffusion Process} We propose a token-independent diffusion process, which independently applies the noise process to each token. We can present the diffusion time steps as: $\mathbf{t}\in \mathbb{R}^{B \times L}$. This approach fundamentally differs from classic sequence-level diffusion models, which use a uniform noise process across all tokens in a sequence ($\mathbf{t}\in \mathbb{R}^{B}$). We denote the token-level diffusion process as: 
\begin{equation} 
x^i_{t_i}= \sqrt{\bar{\alpha}^i_{t_i}}x^i_0 + \sqrt{(1-\bar{\alpha}^i_{t_i})}\epsilon,\quad\epsilon\sim\mathcal{N}(0,I) 
\end{equation}
where $1 \le i \le L$ and $0 \le t \le T$, with $T$ representing the maximum diffusion step. For different indices $i_1$ and $i_2$, the noise process can vary ($t_{i_1} \ne t_{i_2}$). We denote the noisy sequence as $\mathbf{x}_t$ for convenience.

\citet{chenDiffusionForcingNexttoken2024} demonstrated that the token-independent diffusion process optimizes a re-weighting of an Evidence Lower Bound (ELBO) on the expected log-likelihoods $\text{ln} \ p_\theta((\mathbf{x}^i_{t_i})_{1 \le i \le L})$, where the expectation is averaged over noise levels $t_{1:L} \sim [T]^L$. The advantage of the token-level diffusion process stems from the cross-perspective between the diffusion process and masked generative models \cite{liAutoregressiveImageGeneration2024} views noise as a soft mask, interpreting the entire diffusion process as a sequence of soft masking operations (as shown on the left in Figure \ref{fig: overview}). Consequently, even if a few prior tokens are corrupted significantly, the model can still effectively learn to sample from the correct conditional distribution, capturing the distribution of all possible subsequences in the training set. This approach offers potential benefits in terms of model robustness and debiased learning. Furthermore, the token-independent diffusion process provides unique advantages during the inference phase, which we explore in Section \ref{sec: inference}.

\subsection{Training Objective}

In Section \ref{sec: AR strategy}, we adopted the per-step prediction auto-regressive strategy, which necessitates computing the training loss for each output target token. To this end, we use both the recommendation loss (Cross-Entropy, CE, $\mathcal{L}{ce}(\cdot)$) and the denoising loss (Mean Squared Error, MSE, $\mathcal{L}{mse}(\cdot)$) as joint training objectives, which is beneficial for both the recommendation task and embedding distribution modeling. For the recommendation loss, we keep it simple by using full-item CE loss without negative sampling.

We first compute the similarity score $\mathbf{s} \in \mathbb{R}^{B \times L \times N}$, where $N$ is the total number of items in the dataset, by taking the inner product of the reconstructed embedding $\hat{\mathbf{x}}$ and the transpose of the entire item embedding $\mathbf{E} \in \mathbb{R}^{N \times D}$, as follows: \begin{equation} \mathbf{s} = \text{Softmax}(\hat{\mathbf{x}} \odot \mathbf{E}^{\top}) \end{equation} Next, we calculate the total loss $\mathcal{L}$ as: \begin{equation} \mathcal{L} = \mathcal{L}_{ce}(\mathbf{s}, \mathbf{i}_{tgt}) + \mathcal{L}_{mse}(\hat{\mathbf{x}}, \mathbf{x}) \end{equation}

Some works, such as DreamRec, have explored using only the denoising loss as the training objective \cite{yangGenerateWhatYou2023, Yong2024DiffusionRecommendationwithImplicitSequenceInfluence}. However, these approaches often fail to achieve satisfactory results because denoising loss does not directly align with the recommendation task. On the other hand, approaches like DiffuRec have addressed this by retaining recommendation loss as the primary training objective while omitting the denoising loss. However, such a strategy significantly constrains the potential of diffusion models in sequential recommendation.

% \begin{figure}[t]
%     \centering
%     \includegraphics[width=\linewidth]{img/train_stage.pdf}
%     \caption{The three-stage training process and the inference process. In stage 2, the causal attention module is trained from scratch.}
%     \label{fig:stage}
% \end{figure}
% \begin{figure}[t]
%     \centering
%     \includegraphics[width=0.7\linewidth]{img/inference_stage.pdf}
%     \caption{The three-stage training process and the inference process. In stage 2, the causal attention module is trained from scratch.}
%     \label{fig:stage}
% \end{figure}

\begin{table*}[t]

\caption{Main results (\%) on six datasets. The best results are in boldface, and the second-best are underlined. \textbf{Improv.} is the relative improvement of the best method against the second-best one.}

\label{tb: main}
\centering
\resizebox{\textwidth}{!}{%
\begin{tabular}{@{}cccccccccccccc@{}}
\toprule
\textbf{Dataset} & \textbf{Metric} & \textbf{GRU4Rec} & \textbf{BERT4Rec} & \textbf{LightSANs} & \textbf{FEARec} & \textbf{SASRec+} & \textbf{EulerFormer} & \textbf{CORE} & \textbf{SVAE} & \textbf{DreamRec} & \textbf{DiffuRec} & \textbf{ADRec} & \textbf{\color[HTML]{AA0069} Improv.} \\ \midrule
\multirow{2}{*}{Baby} & HR@20 & 5.4576 & 4.0009 & 4.0658 & 3.5786 & 5.5268 & {\ul 5.7906} & 2.7472 & 2.7439 & 0.7648 & 5.382 & \textbf{7.1524} & \textit{\textbf{\color[HTML]{AA0069} 23.52\%}} \\
 & NDCG@20 & 2.2568 & 1.6198 & 1.499 & 1.513 & 2.5197 & 2.4982 & 0.9427 & 1.0086 & 0.2777 & {\ul 2.5649} & \textbf{3.1455} & \textit{\textbf{\color[HTML]{AA0069} 22.64\%}} \\ \midrule
\multirow{2}{*}{Beauty} & HR@20 & 12.6462 & 9.8723 & 9.5874 & 9.7294 & {\ul 15.2038} & 14.7346 & 7.5635 & 3.9204 & 0.6815 & 13.9102 & \textbf{16.8246} & \textit{\textbf{\color[HTML]{AA0069} 10.66\%}} \\
 & NDCG@20 & 5.6347 & 3.9216 & 4.5894 & 4.3869 & {\ul 7.6509} & 7.5428 & 2.6558 & 1.4985 & 0.2728 & 7.3326 & \textbf{8.3214} & \textit{\textbf{\color[HTML]{AA0069} 8.76\%}} \\ \midrule
\multirow{2}{*}{ML-100K} & HR@20 & 18.6858 & 10.1833 & 14.1129 & 9.6959 & 18.4538 & {\ul 19.3524} & 11.8659 & 8.1504 & 3.4395 & 16.0688 & \textbf{22.0699} & \textit{\textbf{\color[HTML]{AA0069} 14.04\%}} \\
 & NDCG@20 & 7.1357 & 3.7504 & 5.0443 & 3.5048 & 7.1034 & {\ul 7.6313} & 4.0904 & 3.2905 & 1.3339 & 6.555 & \textbf{9.0028} & \textit{\textbf{\color[HTML]{AA0069} 17.97\%}} \\ \midrule
\multirow{2}{*}{Sports} & HR@20 & 6.0511 & 4.255 & 4.4246 & 3.9261 & 6.4059 & {\ul 6.4759} & 2.8127 & 2.0019 & 0.7432 & 6.2174 & \textbf{8.1639} & \textit{\textbf{\color[HTML]{AA0069} 26.07\%}} \\
 & NDCG@20 & 2.6069 & 1.7844 & 1.8463 & 1.6689 & {\ul 3.344} & 3.2259 & 0.9802 & 0.8447 & 0.2148 & 3.2067 & \textbf{3.6389} & \textit{\textbf{\color[HTML]{AA0069} 8.82\%}} \\ \midrule
\multirow{2}{*}{Toys} & HR@20 & 5.6216 & 4.9423 & 4.3054 & 6.2075 & 10.7082 & {\ul 10.912} & 6.1481 & 1.5031 & 0.4755 & 9.859 & \textbf{12.0924} & \textit{\textbf{\color[HTML]{AA0069} 10.82\%}} \\
 & NDCG@20 & 2.5605 & 1.9627 & 1.8942 & 3.3492 & 6.0517 & {\ul 6.2564} & 2.217 & 0.6266 & 0.1731 & 5.8508 & \textbf{6.7982} & \textit{\textbf{\color[HTML]{AA0069} 8.66\%}} \\ \midrule
\multirow{2}{*}{Yelp} & HR@20 & 6.3762 & 4.0681 & 3.7219 & 3.005 & 6.6894 & 6.435 & 2.3459 & 1.9238 & 0.7681 & {\ul 6.7315} & \textbf{7.2433} & \textit{\textbf{\color[HTML]{AA0069} 7.60\%}} \\
 & NDCG@20 & 2.5408 & 1.5224 & 1.4837 & 1.0783 & 2.5277 & 2.4252 & 0.8788 & 0.7804 & 0.2477 & {\ul 2.5951} & \textbf{2.8875} & \textit{\textbf{\color[HTML]{AA0069} 11.27\%}} \\ \bottomrule
\end{tabular}%
}
\end{table*}

\subsection{Training Strategy}
Section \ref{sec:intro} mentioned the severe embedding representation collapse issue. Here, we present a three-stage training strategy to mitigate this problem effectively (see Figure \ref{fig:stage}). 

In the first stage, we pre-train the embedding layer using the causal attention module, with the training objective being cross-entropy loss. Our goal is to obtain a semantically rich and structured embedding space. In the second stage, we conduct a 5-epoch warm-up on the backbone of ADRec, freezing the embedding weights. The causal attention module is trained from scratch. The goal is to align the denoising model with the embedding space. Without this alignment, early-stage gradient updates could degrade the pre-trained embedding. Finally, we conduct full-parameter training, leveraging the powerful representation learning capabilities of the diffusion model to optimize the embedding weights further. Experimental results demonstrate that our training method effectively expands the representational capacity of the embedding space and increases the distinction between items. We did not observe any signs of representation collapse in ADRec.

\subsection{Inference Strategy} 
\label{sec: inference}

Existing methods typically employ classical sequence-level diffusion processes, introducing unnecessary noise to the historical sequence during the inference phase. The goal of the denoising model is to reconstruct the clean embedding. Therefore, during the inference phase, noise should only be applied to the last position, while the embeddings at other positions remain clean. However, for the one-step prediction strategy (shown in Appendix Figure \ref{fig: apd step_pre}), during inference, the noised target $\mathbf{x}_t \in \mathbb{R}^{B \times 1 \times D}$ is broadcast across the sequence length dimension to align with the conditional guidance $\mathbf{c} \in \mathbb{R}^{B \times L \times D}$, inadvertently introducing noise into the historical sequence and degrading the model's inference performance. The situation remains problematic for the per-step prediction strategy. The sequence-level diffusion process requires the noise process to remain consistent (shown in the middle of Appendix Figure \ref{fig: apd step_pre}). The uniform noise process $\mathbf{t} = [t, t, \dots, t, t]$ is applied to all target tokens ($\mathbf{x}^{1:L}$) from the maximum diffusion step $T$ down to zero.

A common compromise is to decouple the sequence modeling from the diffusion process. Specifically, a sequence model is first employed to capture the correlations among historical items, and the resulting representations are then passed into a diffusion module, where a non-sequence model, such as an MLP, acts as the denoising model \cite{yangGenerateWhatYou2023, liDimeRecUnifiedFramework2024}. However, relying solely on an MLP for denoising eliminates the opportunity to leverage shared attention mechanisms between the historical sequence and the denoising target. Such mechanisms could provide better guidance for denoising by dynamically directing attention based on historical interactions.

The token-level diffusion process in ADRec provides substantial advantages during the inference phase. Specifically, it allows us to apply pure noise only to the last token position while setting the noise schedule for the other positions to zero (as shown on the right in Figure \ref{fig:stage} and Appendix Figure \ref{fig: apd step_pre}). This approach ensures that, during inference, ADRec can perform diffusion iterations on the target item while still receiving guidance from the clean historical sequence at each time step via attention. The time steps during the inference phase are represented as follows: 
\begin{equation} 
\begin{aligned} 
\mathbf{t} &= [0, 0, \dots, 0, T] \\ \mathbf{x}_t &= [x^1_0, x^2_0, \dots, x^{L-1}_0, x^L_t] 
\end{aligned} 
\label{eq: infer} \end{equation}

\section{Experiments}
\label{sec:exp}
\subsection{Experimental Settings}
\subsubsection{Datasets}
We evaluate the effectiveness of ADRec using six commonly used and publicly available datasets: \textbf{Baby}, \textbf{Beauty}, \textbf{Sports}, \textbf{Toys}, \textbf{ML-100K}, and \textbf{Yelp}. We follow the common preprocessing steps outlined in \cite{liDiffuRecDiffusionModel2023, kang2018self} to ensure a minimum of 5 interactions associated with each user and item. Statistics of the preprocessed datasets are summarized in Appendix Table \ref{tb: apd dataset}.

\subsubsection{Baselines}
To comprehensively demonstrate the effectiveness of ADRec, we select target models from various categories, as follows:
\begin{enumerate}[0]
    \item[$\bullet$] \textbf{RNN-based}: \textbf{GRU4Rec} (ICLR '16) \cite{hidasi2016sessionbasedrecommendationsrecurrentneural}.
    \item[$\bullet$] \textbf{Transformer-based}:
    \textbf{BERT4Rec} \cite{sun2019bert4rec} (CIKM '19); \textbf{LightSANs} \cite{fan2021lighter} (SIGIR '21); \textbf{FEARec} \cite{du2023frequencyenhancedhybridattention} (SIGIR '23); \textbf{SASRec+} \cite{klenitskiy2023turning} (RecSys '23); \textbf{EulerFormer} \cite{tian2024eulerformersequentialuserbehavior} (SIGIR '24).
    \item[$\bullet$] \textbf{Embedding-enhancement:} \textbf{CORE} \cite{hou2022core} (SIGIR 22 short).
    \item[$\bullet$] \textbf{VAE-based}: \textbf{SVAE} \cite{sachdeva2019sequential} (WSDM '19).
    \item[$\bullet$] \textbf{Diffusion-based}: \textbf{DreamRec} \cite{yangGenerateWhatYou2023} (NeurIPS '23); \textbf{DiffuRec} \cite{liDiffuRecDiffusionModel2023} (TOIS '23).
\end{enumerate}
% We adjusted the preprocessing strategy for DiffuRec. See Appendix Section \ref{sec: apd adjusted}.
% Due to space limitations, the implementation details and code repository are provided in Appendix Section \ref{sec: apd implementation}.
% A detailed description of the baseline methods is in Appendix Section \ref{sec: apd baseline}.

\begin{table}[t]
\centering
\caption{Training complexity comparison. $L$ is the sequence length and $d$ is the hidden size.}
\label{tb: complexity}
\resizebox{\linewidth}{!}{%
\begin{tabular}{@{}lcccccc@{}}
\toprule
\textbf{Model} & \textbf{GRU4Rec} & \textbf{SASRec+} & \textbf{DreamRec} & \textbf{DiffuRec} & \textbf{ADRec} \\
\midrule
\textbf{Complexity} & $\mathcal{O}(Ld^2)$ & $\mathcal{O}(Ld^2 + dL^2)$ & $\mathcal{O}(Ld^2 + dL^2)$ & $\mathcal{O}(Ld^2 + dL^2)$ & $\mathcal{O}(Ld^2 + dL^2)$ \\
\bottomrule
\end{tabular}}
\end{table}

\subsubsection{Evaluation Protocol}
Similar to the setup in most existing works, we partition the users into training, validation, and test sets in a ratio of 8:1:1. We evaluate all methods using two widely adopted metrics: Hit Rate (HR) and Normalized Discounted Cumulative Gain (NDCG). HR reflects the model's retrieval ability, while NDCG indicates its ranking performance. In all experiments, we report the results as percentages.

\subsubsection{Implementation Details}

We set the maximum number of training epochs to 500, with a validation interval of 5 epochs. Early stopping is applied if the evaluation metric does not improve on the validation set for four consecutive epochs. The training batch size is fixed at 512; the Adam optimizer has a learning rate of \( 1 \times 10^{-3} \) and a cosine annealing schedule. The embedding size and hidden size are set to 128, and the maximum sequence length is set to 50 for all datasets. We use cross-entropy loss for all items without negative sampling as the recommendation loss for all baselines. For diffusion-based baselines, we use the \textit{truncated linear} schedule for the noise process, with 32 diffusion steps, following \cite{liDiffuRecDiffusionModel2023}. Experimental results show that ADRec is not sensitive to the number of diffusion steps (see Appendix Figure \ref{fig: apd dif_steps}). We evaluated each method over five independent runs and reported the average results.

% Please add the following required packages to your document preamble:
% \usepackage{booktabs}
% \usepackage{multirow}
% \usepackage{graphicx}
% \usepackage[normalem]{ulem}
% \useunder{\uline}{\ul}{}

\subsection{Overall Performance} 

In this section, we evaluate the performance of ADRec by comparing it against ten baseline models across six datasets. The results are presented in Table \ref{tb: main}.

ADRec consistently outperforms all baselines, demonstrating its superior performance. Specifically, compared to the best-performing baselines, ADRec achieves average improvements of 15.45\% on HR@20 and 13.02\% on NDCG@20. These significant improvements underscore the effectiveness of ADRec in capturing three critical aspects: maintaining a structured embedding space without collapse, modeling token interdependency through auto-regression, and jointly modeling per-token distributions through diffusion.

Notably, compared to SASRec+, which closely resembles the pre-training stage of ADRec (stage 1 in Figure \ref{fig:stage}), the superiority of ADRec lies in its distribution modeling capability enabled by the diffusion model. In contrast, DreamRec and DiffuRec suffer from embedding collapse due to inadequate architecture design and training strategies, with performance often falling below that of SASRec+. 
% Furthermore, both models apply the diffusion process only to the last target item. At the same time, ADRec employs a per-token diffusion process, fully leveraging the advantages of diffusion models in sequential recommendation. 

\begin{figure}
    \centering
    \vspace{-0.2cm}
    \includegraphics[width=\linewidth]{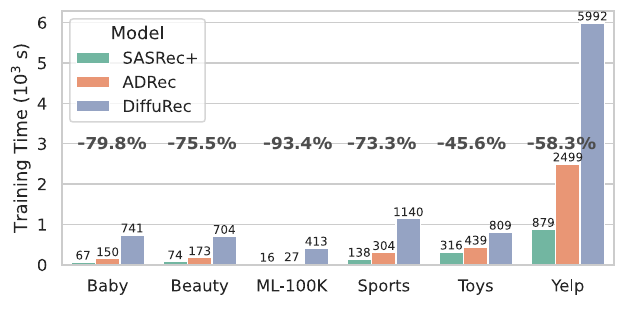}
    \vspace{-0.2cm}
    \caption{Comparison of training time between ADRec and other methods. "-xx\%" indicates the reduction in training time for ADRec compared to DiffuRec.}
    \label{fig: train time}
    \vspace{-0.2cm}
\end{figure}

\subsection{Complexity and Training Time Analysis.} 
The time complexity of ADRec is solely determined by its Transformer backbone, i.e., $\mathcal{O}(nd^2 + n^2d)$, with the diffusion process introducing no additional overhead during training, thus matching the complexity of most baselines, as shown in Table \ref{tb: complexity}. During inference, the time complexity of all diffusion-based methods becomes $\mathcal{O}(T \times (nd^2 + n^2d))$, where $T$ denotes the number of diffusion steps.

For training time, we report the actual GPU wall-clock time from the start of training to early stopping. Stage 1 of ADRec has a duration similar to SASRec+. Stage 2 lasts for only 5 epochs, resulting in a very short training time. Stage 3 requires approximately the same amount of time as Stage 1. As a result, the overall training time of ADRec is approximately twice that of SASRec+.

DiffuRec computes the loss only on the last token of each sequence, sacrificing the parallelism of the Transformer. To cover all interactions, DiffuRec splits a sequence of length $n$ into $n{-}1$ subsequences, where each interaction serves as the last token of a subsequence. This process expands the dataset size by a factor of $n$, leading to an approximate $n$-fold increase in training time per epoch.

Despite using a three-stage training strategy, ADRec reduces total training time by an average of 70.98\% across the six datasets compared to DiffuRec (see Figure \ref{fig: train time}).
\begin{table}[t]
\caption{Results (\%) of ablation experiments conducted for ADRec on six datasets.}
\label{tb: ablation}
\centering
\resizebox{\linewidth}{!}{%
\begin{tabular}{@{}cccccccc@{}}
\toprule
\textbf{Dataset} & \textbf{Metric} & \textbf{ADRec} & \textbf{w/o warm-up} & \textbf{w/o pre-train} & \textbf{w/o mse} & \textbf{w/o CAM} & \textbf{MLP} \\ \midrule
\multirow{2}{*}{\textbf{baby}}    & HR@20   & 7.1524  & 6.7507  & 4.5217 & 4.0439  & 7.0320  & 6.8216  \\
                                  & NDCG@20 & 3.1455  & 2.9535  & 2.1488 & 1.9483  & 3.0853  & 2.8186  \\ \midrule
\multirow{2}{*}{\textbf{beauty}}  & HR@20   & 16.8246 & 15.7483 & 13.233 & 12.6733 & 16.0798 & 16.0738 \\
                                  & NDCG@20 & 8.3214  & 8.1056  & 6.9819 & 6.7209  & 7.9479  & 8.0568  \\ \midrule
\multirow{2}{*}{\textbf{ml-100k}} & HR@20   & 22.0699 & 21.0144 & 14.835 & 13.2913 & 18.9609 & 20.0558 \\
                                  & NDCG@20 & 9.0028  & 8.3206  & 5.5249 & 5.3358  & 7.5569  & 7.8465  \\ \midrule
\multirow{2}{*}{\textbf{sports}}  & HR@20   & 8.1639  & 7.2156  & 5.9635 & 5.6443  & 8.0550  & 7.6604  \\
                                  & NDCG@20 & 3.6389  & 3.7413  & 3.0097 & 2.9709  & 3.5146  & 3.1831  \\ \midrule
\multirow{2}{*}{\textbf{toys}}    & HR@20   & 12.0924 & 11.9226 & 9.3835 & 8.2456  & 12.0264 & 11.4461 \\
                                  & NDCG@20 & 6.7982  & 6.6113  & 5.4178 & 4.8172  & 6.5298  & 6.1759  \\ \midrule
\multirow{2}{*}{\textbf{yelp}}    & HR@20   & 7.2433  & 6.9927  & 7.0224 & 5.6989  & 4.9116  & 6.9563  \\
                                  & NDCG@20 & 2.8875  & 2.8597  & 2.7823 & 2.2123  & 1.9548  & 2.4569  \\ \bottomrule
\end{tabular}%
}
\end{table}

\begin{table*}[ht]
\vspace{-1mm}
\caption{Comprehensive evaluation of embedding representations in the original embedding space.}
\label{tb: Comprehensive evaluation}
\vspace{-2mm}
\centering
\resizebox{\linewidth}{!}{%
\begin{tabular}{@{}cccccccc@{}}
\hline
\textbf{Dataset} & \textbf{Model} & \textbf{Top 5 Singular Values} & \textbf{Singular Value Variance $\uparrow$} & \textbf{Singular Value Entropy $\downarrow$} & \textbf{Covariance Matrix Entropy $\downarrow$} & \textbf{Isotropy Score $\downarrow$} & \textbf{KL Div. to Gaussian $\uparrow$} \\
\hline
\multirow{4}{*}{Baby}
& DreamRec  & [80.58, 80.09, 79.91, 79.02, 78.75]  & 35.29   & 4.85  & 4.83 & 0.51 & 0.96 \\
& DiffuRec  & [106.31, 90.62, 85.69, 81.53, 81.36] & 66.94   & 4.85  & 4.82 & 0.13 & 1.83 \\
& SASRec+   & [84.74, 79.90, 79.44, 78.92, 78.59]  & 38.89   & 4.85  & 4.84 & 0.24 & 1.09 \\
& \textbf{ADRec}     & \textbf{[247.88, 164.00, 145.34, 138.33, 130.06]} & \textbf{1017.59} & \textbf{4.74} & \textbf{4.33} & \textbf{0.02} & \textbf{10.38} \\
\hline
\multirow{4}{*}{Beauty}
& DreamRec  & [89.70, 88.48, 88.32, 87.74, 87.62]   & 33.16   & 4.85  & 4.84 & 0.56 & 0.70 \\
& DiffuRec  & [129.98, 116.93, 113.33, 97.61, 97.35] & 108.99 & 4.84  & 4.81 & 0.10 & 2.25 \\
& SASRec+   & [128.70, 125.14, 106.41, 97.86, 94.20] & 101.55 & 4.84  & 4.81 & 0.09 & 2.11 \\
& \textbf{ADRec}     & \textbf{[177.19, 159.37, 145.52, 137.74, 126.75]} & \textbf{995.55} & \textbf{4.76} & \textbf{4.51} & \textbf{0.02} & \textbf{10.37} \\
\hline
\multirow{4}{*}{ML-100K}
& DreamRec  & [43.16, 42.64, 42.42, 41.83, 41.47]    & 33.56   & 4.84  & 4.78 & 0.23 & 4.41 \\
& DiffuRec  & [55.42, 49.44, 44.24, 42.55, 41.41]    & 38.75   & 4.83  & 4.77 & 0.04 & 5.15 \\
& SASRec+   & [47.37, 42.03, 41.78, 41.60, 41.50]    & 39.33   & 4.83  & 4.78 & 0.11 & 4.60 \\
& \textbf{ADRec}     & \textbf{[90.91, 82.50, 71.20, 63.54, 53.68]}     & \textbf{122.57} & \textbf{4.80} & \textbf{4.55} & \textbf{0.04} & \textbf{10.43} \\
\hline
\multirow{4}{*}{Sports}
& DreamRec  & [122.18, 121.81, 121.40, 121.04, 120.88] & 33.16  & 4.85  & 4.85 & 0.66 & 0.35 \\
& DiffuRec  & [220.33, 182.85, 172.72, 151.13, 137.70] & 347.81 & 4.84  & 4.79 & 0.09 & 3.43 \\
& SASRec+   & [133.91, 125.74, 122.40, 122.00, 120.98] & 68.15  & 4.85  & 4.84 & 0.13 & 0.81 \\
& \textbf{ADRec}     & \textbf{[415.16, 355.94, 317.40, 275.30, 257.72]} & \textbf{4333.00} & \textbf{4.65} & \textbf{3.97} & \textbf{0.01} & \textbf{10.37} \\
\hline
\multirow{4}{*}{Toys}
& DreamRec  & [109.57, 107.46, 106.19, 104.88, 103.48] & 91.02   & 4.85  & 4.82 & 0.40 & 1.59 \\
& DiffuRec  & [139.38, 135.46, 117.65, 106.74, 105.77] & 138.12  & 4.84  & 4.81 & 0.10 & 2.39 \\
& SASRec+   & [103.28, 97.70, 96.78, 96.15, 95.16]     & 52.63   & 4.85  & 4.84 & 0.13 & 1.03 \\
& \textbf{ADRec}     & \textbf{[182.03, 156.40, 148.03, 139.77, 137.68]} & \textbf{1243.76} & \textbf{4.75} & \textbf{4.51} & \textbf{0.01} & \textbf{10.37} \\
\hline
\multirow{4}{*}{Yelp}
& DreamRec  & [265.59, 265.08, 264.78, 264.31, 264.19] & 32.51   & 4.85  & 4.85 & 0.84 & 0.06 \\
& DiffuRec  & [772.81, 690.32, 623.17, 587.23, 577.26] & 13798.54 & 4.75 & 4.27 & 0.03 & 10.36 \\
& SASRec+   & [843.78, 675.20, 636.04, 625.65, 568.57] & 12385.48 & 4.77 & 4.30 & 0.03 & 10.36 \\
& \textbf{ADRec}     & \textbf{[810.29, 781.18, 625.42, 566.81, 541.51]} & \textbf{18955.47} & \textbf{4.68} & \textbf{4.12} & \textbf{0.01} & \textbf{10.36} \\
\hline
\end{tabular}
}
\vspace{-3mm}
\end{table*}

\subsection{Ablation Study}
To evaluate the effectiveness of each design choice in ADRec, we perform ablation studies with four variants:

\begin{enumerate}[0] \item[-] (Variant 1) \textbf{w/o warm-up}: ADRec without warmup stage (training stage 2). \item[-] (Variant 2) \textbf{w/o pre-train}: ADRec trained in an end-to-end manner without pre-training. \item[-] (Variant 3) \textbf{w/o MSE}: End-to-end ADRec without the denoising loss $\mathcal{L}_{mse}(\hat{\mathbf{x}}, \mathbf{x})$. \item[-] (Variant 4) \textbf{w/o CAM}: ADRec removed CAM module. \item[-] (Variant 5) \textbf{MLP}: ADRec with an MLP as the denoising model. \end{enumerate}

Table \ref{tb: ablation} presents the ablation results across six datasets. Comparing the original ADRec with Variant 1 reveals that aligning pre-trained embeddings with the model backbone is crucial for performance. The significant performance drop, compared to Variant 2, emphasizes the importance of the three-stage training strategy, which provides a meaningful and structured embedding space to the diffusion module and mitigates embedding collapse. In contrast to Variant 3, we find that jointly optimizing the denoising loss and recommendation loss is essential for effectively deploying the diffusion model, leading to improved recommendation performance. Without CAM, compared with Variant 4, ADM can still extract historical interaction information through self-attention, but a performance drop is observed, especially on ML-100K and Yelp. Finally, when comparing with Variant 5, which employs an MLP for denoising, we observe that even with the same MLP architecture used in DreamRec, ADRec's careful architectural design and training strategy still ensure superior performance. In comparison, DreamRec's performance is nearly equivalent to random recommendation, underscoring the effectiveness of our approach.

\begin{table}[t]
\caption{Linear Probe accuracy of different models on ML-100K. Each item is multi-class, with a total of 26 classes.}
\label{tb: linearprobe}
\centering
% \resizebox{0.9\linewidth}{!}{%
\begin{tabular}{@{}ccccc@{}}
\toprule
\textbf{} & DreamRec & DiffuRec & SASRec+ & ADRec \\ \midrule
Precision & 20\% & 35\% & 42\% & \textbf{44\%} \\
Recall & 14\% & 30\% & 40\% & \textbf{46\%} \\
F1-score & 16\% & 31\% & 40\% & \textbf{46\%} \\ \bottomrule
\end{tabular}%
\end{table}

\subsection{Embedding Collapse}
\subsubsection{t-SNE Visualization}
Recent studies \cite{fuest2024diffusionmodelsrepresentationlearning} have highlighted that diffusion models excel not only in continuous generation tasks but also in representation learning. To illustrate that ADRec can learn a more structured representation space, we visualize the embedding weights of ADRec, DiffuRec, DreamRec, and SASRec+ (the latter is also considered as ADRec in its pre-training stage) using T-SNE \cite{van2008visualizing}, as shown in Figure \ref{fig:tsne} and Appendix Figure \ref{fig:tnse_full}.

We begin by focusing on the shapes of the visualizations. The embeddings of DiffuRec and DreamRec closely resemble isotropic standard Gaussian noise, particularly in their contour shapes, which indicate embedding collapse. While ADRec retains distinct structured features. More importantly, the scale of the representation space expanded clearly in ADRec. As indicated by the scale label in the top-left corner of the figure, ADRec effectively enlarges the weak representation space in the pre-train stage (similar to SASRec+), making the embeddings more distinguishable.

\subsubsection{Comprehensive Evaluation of Embedding Collapse in Original Embedding Space.}
Beyond the t-SNE visualizations, we provide additional quantitative analyses in the original embedding space, as presented in Table \ref{tb: Comprehensive evaluation}:
\begin{enumerate}[0]
    \item[-] Top-5 singular values and singular value variance are much higher in ADRec, indicating strong principal component dominance and the ability to capture more discriminative features. Other models (especially DreamRec) show flatter singular values, implying weak feature directions.
    
    \item[-] Singular value entropy, covariance entropy, and isotropy score are lower for ADRec, demonstrating more directional (anisotropic) embeddings and better parameter efficiency. This highlights another aspect of ADRec's superior performance: ADRec does not just have more information; it has information concentrated in effective directions.
    
    \item[-] Silhouette scores are also higher for ADRec, suggesting clearer clustering among item embeddings.
\end{enumerate}
\textit{All these metrics indicate that the characteristics of the original embedding space for all DM based models align closely with the visualizations produced by t-SNE.} DreamRec still resembles a Gaussian distribution, similar to randomly initialized embeddings, while ADRec exhibits the most anisotropy.

\begin{figure}
    \centering
    
    \includegraphics[width=\linewidth]{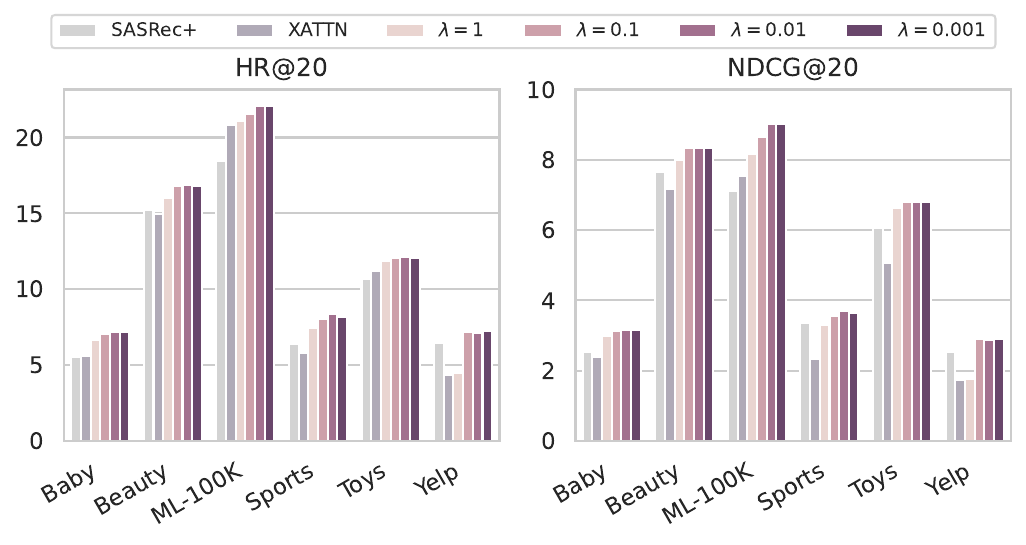}
    \vspace{-4mm}
    \caption{Comparison of ADRec with linear integration using different $\lambda$ coefficients and cross-attention integration, with SASRec+ as a baseline.}
    \label{fig: lambda}
    \vspace{-2mm}
\end{figure}

\subsubsection{Linear Probing}
We also conduct a linear probing experiment to further confirm the importance of the expanded and structured embedding space in ADRec. The experimental details are provided in Appendix Section \ref{sec: apd linear prob}. From the results in Table \ref{tb: linearprobe}, it is clear that ADRec achieves the best performance across all three metrics, whereas DiffuRec and DreamRec suffer from poor performance due to embedding collapse. Compared to SASRec+, the expanded embedding space of ADRec enhances the model’s ability to recognize embeddings, thereby improving its overall performance.

\subsection{Feature Aggregation Method}

In this section, we explore various feature aggregation methods. Specifically, as illustrated in Figure \ref{fig: lambda}, we investigate the impact of varying $\lambda$ coefficients in the linear integration method, following the approach used in DiffuRec. Our results indicate that ADRec demonstrates considerable robustness to changes in $\lambda$, with most settings outperforming SASRec+. A moderate enhancement in the efficacy of recommendations is noted as $\lambda$ diminishes. 

Notably, the training strategy of ADRec enhances the robustness of the embedding space, enabling the sequence model to tolerate greater noise introduced into the historical sequence without causing significant degradation of the embeddings. In contrast, DiffuRec is highly sensitive to $\lambda$, as highlighted in the original paper. For example, when $\lambda = 0.1$, performance declines by up to 80\% compared to $\lambda = 0.001$, which worsens the collapse of DiffuRec's weak embedding space. Furthermore, we experimented with feature integration using cross-attention; however, this approach resulted in decreased performance.

\begin{figure}
    % \vspace{-2mm}
    \centering
    \includegraphics[width=\linewidth]{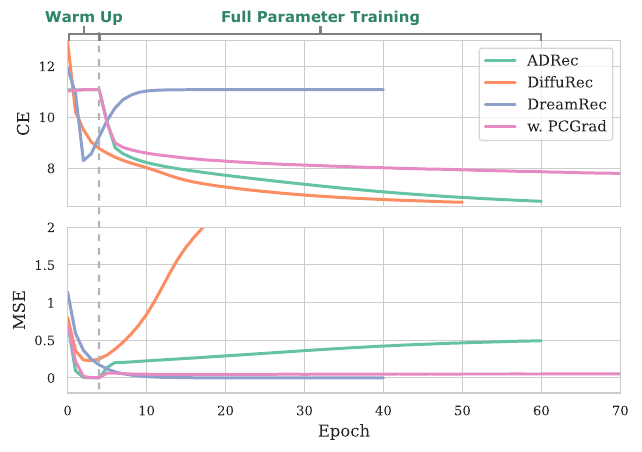}
    % \vspace{-4mm}
    \caption{Training loss curves on the Yelp dataset. “w. PCGrad” denotes ADRec trained using the PCGrad framework. The pre-training stage of ADRec is omitted for clarity.}

    \label{fig: apd loss}
    % \vspace{-2mm}
\end{figure}

\subsection{Joint Optimization Objective of ADRec}
\label{sec: pcgrad}
Diffusion-based recommendation models frequently encounter the challenge of selecting the appropriate training objective. 
 
Figure \ref{fig: apd loss} illustrates the training loss curve on Yelp. We observe that DiffuRec and DreamRec, which focus on a single objective, experience degradation in the other objective, adversely affecting overall recommendation performance. In contrast, for ADRec, we notice that losses display a distinct step-wise pattern throughout training. During the backbone warm-up stage, the convergence for the recommendation loss (CE) and denoising loss indicates alignment between the backbone network and the pre-trained embedding weights. In the full-parameter training phase, the recommendation loss continues to decline, suggesting further optimization of the embedding weights. Although the denoising loss shows a slight upward trend, it remains minimal compared to DiffuRec.

\citet{liDimeRecUnifiedFramework2024} argues that a joint training objective of CE and MSE may lead to potential gradient conflicts. To verify this, we incorporate the PCGrad technique \cite{yu2020gradient} into ADRec, which projects conflicting gradients to maintain an acute angle between their directions during optimization. 

\begin{figure}
    \centering
    % \vspace{-2mm}
    \includegraphics[width=\linewidth]{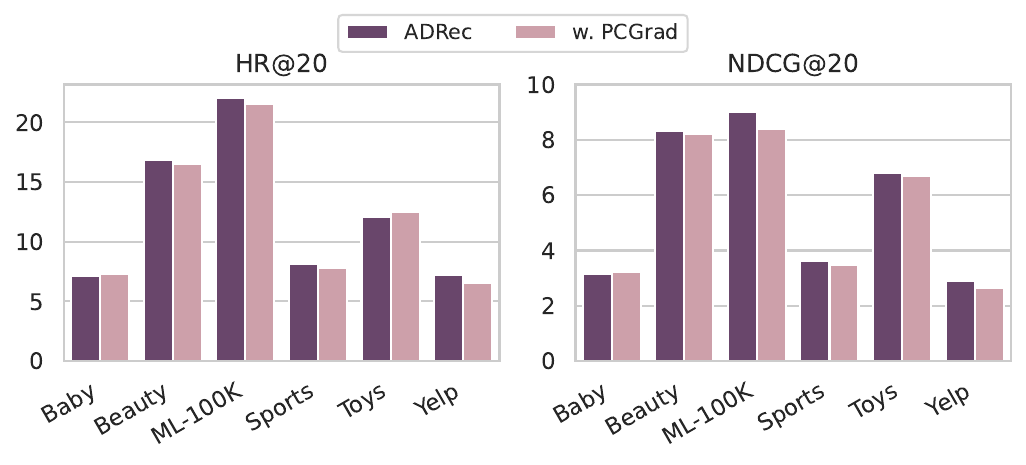}
    % \vspace{-4mm}
    \caption{Performance impact of introducing the PCGrad framework on six datasets.}
    \label{fig: pcgrad}
    % \vspace{-4mm}
\end{figure}

The joint optimization objective can be viewed as a trade-off between the generation and classification tasks. After incorporating the PCGrad framework, the degradation in denoising loss is notably suppressed; however, the recommendation loss does not reach the original ADRec level. And a slight degradation in recommendation performance is observed in Figure~\ref{fig: pcgrad}. In contrast to prior approaches, ADRec demonstrates only mild gradient conflict. While its optimization leans toward the recommendation objective, it maintains a meaningful level of denoising learning—unlike DiffuRec, which effectively neglects it. This balanced trade-off plays a key role in enhancing recommendation quality.

% \subsection{Other Results}

% In Appendix Section \ref{sec: apd PE}, we explain the rationale behind ADRec's decision to omit position encoding. By employing causal masking, the attention mechanism can approximate the absolute position of tokens based on the number of observable tokens. In Appendix Section \ref{sec: apd dif steps}, we discuss ADRec’s insensitivity to the number of diffusion time steps, attributing this characteristic to auto-regression, which serves as a performance lower bound. In Appendix Section \ref{sec: apd pcgrad}, we examine the potential optimization conflict between the recommendation loss and the denoising loss. Our analysis shows that ADRec primarily prioritizes the recommendation objective, while ensuring the denoising target does not degrade significantly.

\section{Conclusion}
In this paper, we identify and address the embedding collapse problem in diffusion-based sequential recommendation models. We propose ADRec, a novel framework that integrates an auto-regressive strategy with a token-level diffusion process. By decoupling embedding optimization into semantic pretraining and subsequent denoising, the proposed three-stage training strategy effectively prevents embedding collapse that can occur when denoising is applied directly to randomly initialized embeddings. By tackling key challenges such as weak representation spaces and embedding collapse, ADRec significantly improves the model’s capacity to learn meaningful item distributions and sequence dynamics.

% While our results demonstrate significant advancements, there remains considerable potential for further improvements. We believe that ADRec lays a solid foundation for the diffusion-based sequential recommendation methods.

While our results demonstrate significant advancements, there remains considerable potential for further improvement in utilizing diffusion models for recommendation systems. ADRec provides a solid foundation for diffusion-based sequential recommendation methods, with opportunities for future enhancements.

\begin{acks}
\sloppy
This work is supported by the Natural Science Foundation of China No.~62472196, Jilin Science and Technology Research Project 202301 01067JC, National Key R\&D Program of China under Grant No.~2021 ZD0112501 and 2021ZD0112502, National Natural Science Foundation of China under Grant No.~62272193, National Key R\&D Program of China under Grant Nos.~2022YFB3103700 and 2022YFB3103702.
\fussy
\end{acks}

%%% -*-BibTeX-*-
%%% Do NOT edit. File created by BibTeX with style
%%% ACM-Reference-Format-Journals [18-Jan-2012].

%% The next two lines define the bibliography style to be used, and
%% the bibliography file.
% \bibliographystyle{unsrt}
% \bibliographystyle{ACM-Reference-Format}
% % \balance
% \bibliography{sample-base}

%%
%% If your work has an appendix, this is the place to put it.
% \newpage
\appendix

\section{Related Works}
\label{sec: apd related work}

\subsection{Traditional Sequential Recommendation}
Recent sequential recommendation (SR) advancements have leveraged powerful deep neural networks to capture item interdependencies. GRU4Rec \cite{hidasi2016sessionbasedrecommendationsrecurrentneural} employs a gated recurrent unit (GRU) to model the temporal dependencies within user sequences. SASRec \cite{kang2018self} adopts a multi-layer transformer architecture to efficiently model sequence interactions. SASRec+ \cite{klenitskiy2023turning}, an enhancement of SASRec, replaces the original binary cross-entropy loss with cross-entropy loss, significantly improving performance. BERT4Rec \cite{sun2019bert4rec} extends this approach by incorporating bidirectional transformer layers and using a Cloze task to capture user sequence patterns better. LightSANs \cite{fan2021lighter} introduces a low-rank decomposition to the self-attention mechanism, enhancing both the efficiency and effectiveness of transformer-based SR models. EulerFormer \cite{tian2024eulerformersequentialuserbehavior} introduces a unified theoretical framework to capture both semantic and positional differences between items in a transformer, enhancing the model's expressive power in sequence modeling. These methods have demonstrated strong sequence modeling capabilities and provide valuable insights for our work.

\subsection{Diffusion Models for SR}

In recent years, diffusion models have been increasingly applied to sequential recommendation, aiming to improve the quality of the embedding space. DiffRec \cite{wangDiffusionRecommenderModel2023} falls within the collaborative recommendation domain, where it predicts whether a user will interact with each item. DreamRec \cite{yangGenerateWhatYou2023} employs a diffusion model to explore the underlying distribution of target items, generating an oracle next-item embedding that aligns with user preferences, eliminating the need for negative sampling. DiffRIS \cite{Yong2024DiffusionRecommendationwithImplicitSequenceInfluence} enhances behavior sequence-based guidance representations by explicitly modeling both long- and short-term user interests. However, DiffRec and DreamRec rely solely on denoising loss, which does not align with the recommendation task. DiffuRec \cite{liDiffuRecDiffusionModel2023}, the most substantial open-source baseline, performs denoising only on the final target item, using cross-entropy loss. This approach limits the model's capacity to capture sequence dynamics and item distribution. DimeRec \cite{liDimeRecUnifiedFramework2024} introduces joint training by applying both denoising loss and recommendation loss to the final output item and mitigates gradient conflicts in the joint loss by introducing noise to the final target item embedding using a complex geodesic random walk. However, existing diffusion-based methods rarely address the issue of embedding collapse.
\begin{figure}[t]
    \centering
    \includegraphics[width=\linewidth]{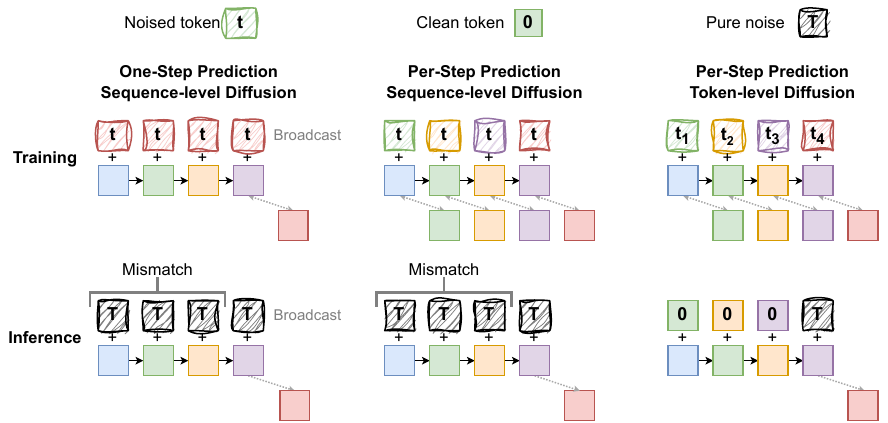}
    \caption{Diagrams of different auto-regressive strategies and diffusion strategies.}
    \label{fig: apd step_pre}
\end{figure}

\begin{table}[t]
\caption{Statistics of datasets after preprocessing.}
\label{tb: apd dataset}
\centering
\resizebox{\linewidth}{!}{%
\begin{tabular}{@{}lcccccc@{}}
\toprule
             & Baby    & Beauty  & ML-100K & Sports  & Toys    & Yelp    \\ \midrule
Users        & 11761   & 10553   & 938     & 22686   & 11268   & 136346  \\
Items        & 4731    & 6086    & 1008    & 12301   & 7309    & 64669   \\
Interactions & 92613   & 94119   & 54457   & 185779  & 95468   & 1857033 \\
Avg. length  & 7.89    & 8.92    & 58.01   & 8.19    & 8.47    & 13.62   \\
Sparsity     & 99.62\% & 99.74\% & 94.50\% & 99.63\% & 99.95\% & 99.98\% \\ \bottomrule
\end{tabular}%
}
\end{table}

\begin{table}[t]
\caption{The impact of positional encoding on ADRec performance. "with PE" means ADRec with positional encoding.}
\label{fig: PE}
\centering
\resizebox{\linewidth}{!}{%
\begin{tabular}{@{}cccccccc@{}}
\toprule
Model                    & Metric  & Baby   & Beauty  & ML-100K & Sports & Toys    & Yelp   \\ \midrule
\multirow{2}{*}{ADRec}   & HR@20   & \textbf{7.1524} & \textbf{16.8246} & \textbf{22.0699} & \textbf{8.1639} & \textbf{12.0924} & 7.2433 \\
                         & NDCG@20 & \textbf{3.1455} & \textbf{8.3214}  & \textbf{9.0028}  & \textbf{3.6389} & \textbf{6.7982}  & 2.8875 \\ \midrule
\multirow{2}{*}{with PE} & HR@20   & 7.0034 & 16.3726 & 21.0172 & 7.9949 & 11.5668 & \textbf{7.5212} \\
                         & NDCG@20 & 2.9843 & 7.8961  & 8.4448  & 3.2769 & 5.8444  & \textbf{3.1355} \\ \bottomrule
\end{tabular}%
}
\end{table}

\balance

\section{Adjusted preprocessing strategy for DiffuRec}
\label{sec: apd adjusted}

DiffuRec originally adopted a subsequence splitting strategy that may \textit{unfairly} leverage more training data. For instance, given a sequence of length 200 with a maximum truncation length of 50, standard preprocessing keeps only the last 50 interactions. In contrast, DiffuRec splits all 200 interactions into training subsequences. \textit{To ensure a fair comparison with other baselines, we revised its preprocessing pipeline: sequences are first truncated to the maximum length and only then split into subsequences.}

% \balance
\section{Linear Probe}
\label{sec: apd linear prob}
To verify whether the model has formed a structured embedding representation, we conducted a linear probing experiment on the ML-100K dataset. We first load the trained embedding weights, perform a batch normalization (BatchNorm1d) operation, and then feed them into a linear classification head. During training, the embedding weights are frozen. In the ML-100K dataset, each item has 26 category attributes:
['Action', 'Crime', 'Film-Noir', 'Musical', 'Sci-Fi', 'Adventure', 'Animation', 'Biography', 'Comedy', 'Documentary', 'Drama', 'Family', 'Fantasy', 'Game-Show', 'History', 'Horror', 'Music', 'Mystery', 'News', 'Reality-TV', 'Romance', 'Short', 'Sport', 'Talk-Show', 'Thriller', 'War', 'Western']. We use a BCE loss function, with an Adam optimizer and a learning rate of 0.01, and train for 20 epochs.

\section{Positional Encoding}
\label{sec: apd PE}
We find that ADRec does not require positional encoding. Adding positional encoding leads to a slight decrease in performance (Figure \ref{fig: PE} in the Appendix). This is similar to the conclusion in \cite{haviv2022transformer}, where it was observed that in some tasks, causal attention allows the model to approximate the absolute position of each token by inferring the number of preceding tokens it can attend to. This is closely related to the per-step prediction auto-regressive strategy we employ.

It becomes more complicated when introducing diffusion models. Potential insights based on our understanding are as follows: 
\begin{enumerate}[0]
\item[-] The noisy input at each diffusion step may already be disrupted, rendering the positional encoding information unstable or meaningless.
\item[-] Forcing positional encoding may lead the Transformer to learn “pseudo-patterns,” introducing additional bias and interfering with the true denoising process.
\item[-] Positional encoding also acts as meaningless noise for the diffusion model, deviating from the assumption that noise gradually converges to a Gaussian distribution.

\end{enumerate}

\begin{figure}[t]
    \centering
    \includegraphics[width=\linewidth]{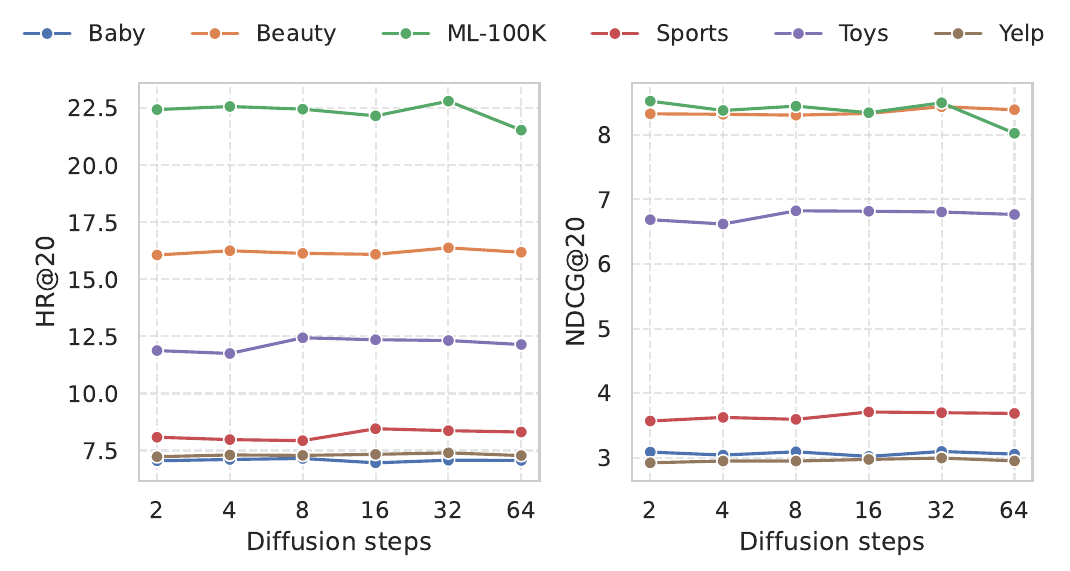}
    \caption{Recall@20 and NDCG@20 at different diffusion steps}
    \label{fig: apd dif_steps}
\end{figure}

\section{Diffusion Steps}
\label{sec: apd dif steps}
In Appendix Figure \ref{fig: apd dif_steps}, we demonstrate that ADRec is not sensitive to the diffusion time steps. This could be because the auto-regressive and diffusion models jointly drive the recommendation results. The discriminative power of the auto-regressive model over the structured embedding space defines the lower bound of ADRec's performance, while a substantial embedding space relies on the distribution modeling capabilities of the diffusion model.

\begin{figure*}[t]
    \centering
    \includegraphics[width=0.99\linewidth]{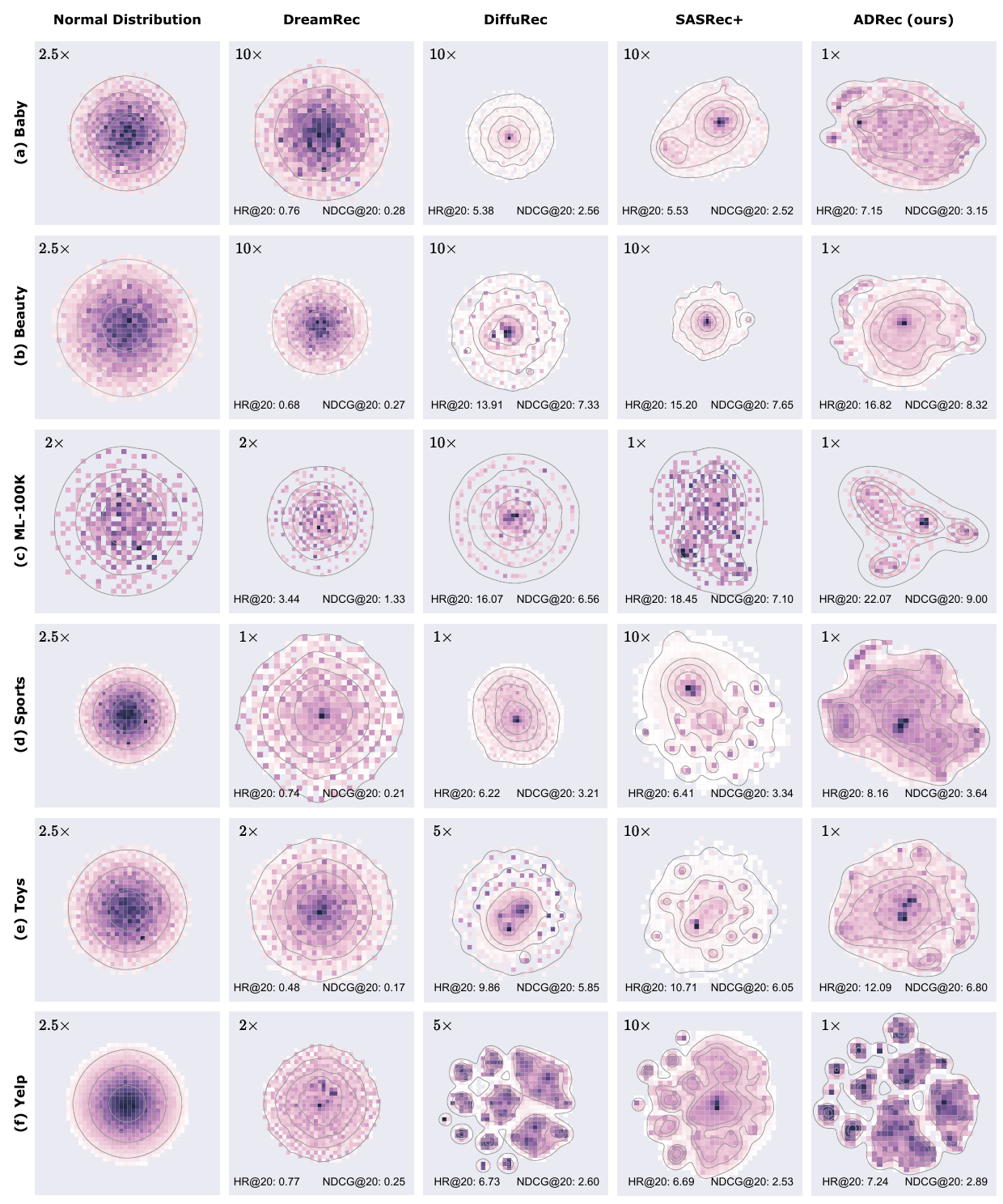}
    % \vspace{-0.4cm} 
    \caption{t-SNE results on six datasets.}
    \label{fig:tnse_full}
\end{figure*}

\end{document}